\documentstyle[prd,aps,floats,graphics,epsfig]{revtex}

\begin{document}


%
%
\input epsf

\renewcommand{\topfraction}{0.99}
\renewcommand{\bottomfraction}{0.99}
\newcommand{\pbar}{\not{\!\partial}}
\newcommand{\dbar}{\not{\!{\!D}}}
\twocolumn[\hsize\textwidth\columnwidth\hsize\csname
@twocolumnfalse\endcsname

\vspace{-1cm}

\begin{flushright}
SHEP/00/01
\end{flushright}

\vspace{1cm}

\title{Gravitino production in hybrid inflationary models}
\author{Mar Bastero-Gil $^{\dagger}$ and Anupam Mazumdar $^{\ast}$ }
\address{$ ^{\dagger}$ Department of Physics, University of Southampton, 
Southampton, SO17 1B JW, United Kingdom \\
$^{\ast}$ Astrophysics Group, The Blackett Laboratory, Imperial College,
London, SW7 2BZ, United Kingdom.
}
\maketitle
\begin{abstract}
It has been recently shown that it is possible to excite gravitinos in an
expanding background due to time-varying scalar field oscillating at the
bottom of the inflationary potential. The two components of the gravitino,
namely helicity $1/2$ and helicity $3/2$ are excited differently due to
the presence of different time-varying mass scales in the problem. In this
paper we analyse the production of both the helicities in multi-chiral
scenario, in particular focusing on a general model of hybrid inflation. 
Fermion production in hybrid models is very much different from that of the
chaotic models discussed so far in the literature. In this paper we give a
full account of gravitino production analytically and numerically.  It is
noticed that the creation of gravitinos does not take place in the first few
oscillations of the inflaton field, rather the production is a gradual
and delayed process. It takes roughly $30-40$ oscillations to build up the
production and for the saturation to take place it can even take longer time,
depending on the model parameters.  We give an estimation of the
reheat temperature and a brief discussion upon back-reaction on the
gravitino production which could change its abundance.

\end{abstract}


\vskip2pc]

\section{Introduction}
Low energy effective $N=1$ supergravity is a predictive theory
\cite{nilles}, which could predict an inflationary potential flat
enough to provide adequate density perturbations \cite{lyth}. So far, such
viable inflationary models were constrained from observations by fixing the
height of the potential, which essentially determines the amplitude of the
COBE normalization. The first and the second derivative of the potential
determines the tilt in the power spectrum, and the Yukawa couplings
of the inflaton to other particles determines the reheat temperature of the 
Universe. The higher the coupling constant is, the higher is the temperature 
of the thermal bath and so the creation of the gravitinos from the collisions 
or the decay of other particles. The gravitinos decay very late and depending 
on their mass their life time could be long enough to disrupt the synthesis 
of light elements, via hadronic shower or by altering the entropy 
density of the baryons. Due to these reasons there is a strong constraint
on the reheat temperature, for a review see Ref.~\cite{sarkar}.

However, there is also a non-thermal phase of the Universe, just after the end of
slow-roll inflation when the scalar field begins to oscillate coherently at
the bottom of the potential. During this era an explosive production of
particles, both bosons \cite {linde} and fermions \cite{patrick,peloso},
may take place due to the non-perturbative decay of the inflaton to other
fields. It has also been shown that it is possible to create
super-massive bosons and fermions. However, fermionic production is
always saturated by the Pauli blocking. As a matter of fact creation of heavy
non-thermal bosons can be a good candidate for weakly interacting massive
particles, known as WIMPS \cite{rocky}. The decay of super-massive bosons can 
explain the ultra-high energy cosmic rays \cite{berezin}.
Super-massive fermions can be used in leptogenesis, mainly from the decay
of right-handed neutrinos into Higgs and leptons, which explicitly violates
the CP conserving phase \cite{peloso}. It is worth mentioning that the
non-perturbative technique of decaying inflaton to other particles has
given a new paradigm shift in understanding the hot big bang universe from
the ultra-cold inflationary regime.

Recently preheating in the context of global supersymmetric theories has
been considered \cite{mar,mar1} and, as a natural extension it has been
necessary to consider a local version of the supersymmetric theory and discuss
the non-perturbative aspects of particle production and their consequences
to nucleosynthesis. The local version of supersymmetry, known as
supergravity, naturally accommodates the graviton and its superpartner the 
gravitino, a spin $3/2$ particle. Quantization of spin $3/2$ particles
in the presence of an external background is plagued by consistency
problems.  It has been known for a long time that quantization of spin
$3/2$ particles in scalar, electromagnetic, or gravitational backgrounds
can give rise to acausal behaviour \cite{velo}. Supergravity is the only
set-up where such problems do not occur, provided the background fields
also satisfy the corresponding equations of motion \cite{deser0}. 
Nevertheless, the complicated form of the Rarita-Schwinger
equation makes it extremely difficult to extract any explicit results even
in a simple background \cite{auvil}. The problem was first addressed in
Ref.~\cite{deser}, where the authors quantized spin $3/2$ particles in a
non-vanishing cosmological background, almost two decades ago.

The slightest generalization of quantizing spin $3/2$ has been done very
recently in literature, in Ref.~\cite{anupam}. The authors have extended
the calculation of quantizing spin $3/2$ in presence of a time-varying
homogeneously oscillating scalar field in a cosmological background.  This
is the first result where the non-perturbative decay of the inflaton to
gravitinos during preheating has been taken into account.  The authors have
explicitly shown the production of a particular helicity, the $\pm 3/2$
components of gravitino, in a particular new-inflationary type model
\cite{ross}. It has been 
noticed that the non-perturbative result can give rise to a larger 
abundance compared to the perturbative decay.
The gravitino to photon number density has been found to be 
$n_{3/2}/n_\gamma \sim 10^{-12}$ \cite{anupam}.  This abundance is 
$3$ orders of magnitude larger than the
thermal abundance for a reheat temperature $10^{5}$ GeV \cite{sarkar}.
Such over production of gravitinos has been the first proof of the
non-thermal production of gravitinos with helicity $\pm 3/2$, which demands
constraining the reheat temperature in any supergravity motivated
inflationary model.

However, a massive gravitino has $4$ degrees of freedom, the
other two degrees of freedom due to the helicity $\pm 1/2$ components
of the gravitino. The 
production of helicity $\pm 1/2$ gravitinos is directly related to the problem
of super-Higgs mechanism in supergravity models, which was studied in
Ref.~\cite{deser} in the context of a non-vanishing cosmological constant,
and in Ref.~\cite{cremmer} with a vanishing cosmological constant.  
In the presence of a
time-varying scalar field there is an additional source of supersymmetry
breaking via the non-vanishing time derivative of the homogeneous scalar
field. This plays an important role in the context of cosmology when the
scalar field is recognized as the inflaton, oscillating coherently at the
bottom of the scalar potential. Due to the presence of such a field,
supersymmetry is always broken at the minima of the potential and
the initially massless gravitino which possesses only the helicity $\pm 3/2$
components ``eats'' the goldstino to gain the other $\pm 1/2$
components. 
At this
point, one may wonder how to generalize the super-Higgs mechanism in
such a scenario. In fact the problem turns out to be
quite complicated and it has been addressed in two seminal papers
\cite{kallosh,riotto1}. These papers also study for the first time the
production mechanism of the helicity $\pm 1/2$ components of gravitino ( see
other papers in the similar context \cite{lyth1}). In all these
papers the authors have assumed the existence of a unitary gauge, where 
the physical Lagrangian is free from the goldstino field. We 
will explicitly show that it is possible to choose such a gauge, where the 
gravitino equation of motion is free from the goldstino field. Our 
calculation is valid for more than one chiral field as well. Though we 
shall give a proof for the $F$-type supersymmetry breaking, this can be
extended to $D$-type supersymmetry breaking also.

It has been noticed in Refs.\cite{kallosh,riotto1} that the production of
the two helicity states
are completely different. The helicity $\pm 1/2$ components are
produced copiously compared to the helicity $\pm 3/2$. In the case of
helicity $\pm 3/2$, conformal 
invariance is broken due to the presence of a time-varying gravitino mass,
which is 
usually Planck mass suppressed, whereas for helicity $\pm 1/2$  the breaking of
conformal invariance is related to the presence of a massive Goldstone 
fermion \cite{kallosh,riotto1}, whose time-varying mass is not 
suppressed by the Planck mass. Moreover in the high momentum limit the 
helicity $\pm 1/2$ gravitino behaves like a fermion, the goldstino, as it 
is stated by the  equivalence principle \cite{riotto2,maroto}. 
This has been studied in a single-chiral field scenario, where the  
source of conformal breaking can be directly related to the mass of
Goldstino. Unfortunately in the multi-chiral field scenario the
quantization scheme becomes more involved, and the relation between
conformal breaking and Goldstino mass is not so straightforward. The
situation has been briefly discussed in Ref. \cite{riotto2}, and an
attempt of a perturbative scheme has been suggested. However, it would
be nice to discuss a non-perturbative scheme. 
Our paper fills that gap and, as we shall see, we can  discuss
non-perturbative production of the helicity $\pm 1/2$ gravitinos in
the multi-chiral case.

The best example to study the multi-chiral field scenario is in the context 
of a general class of supersymmetric hybrid inflation model \cite{mar,mar1}.
There are essentially two scalar fields, one is responsible for inflation,
and the other field is responsible for the phase transition which
results in terminating  
the inflationary era. Unlike the non-supersymmetric version of hybrid 
inflation model \cite{hybrid}, the supersymmetric version considered
in this paper  
has only one coupling constant in the potential. This leads to  
a single natural frequency of oscillations. This
gives us an ample opportunity to use techniques to explore 
gravitino production in a similar spirit as in the case of a single-chiral
field. Fermionic production in this case 
is very much different compared to that of a quadratic inflationary
potential.  
In the hybrid models the effective mass term for the fermions is
always positive and as a 
result the production can never be completed in just a few
oscillations, rather the occupation number gradually increases and depending
on the model parameters the rate of production could be slowed down
significantly. For example,  
in order to reach the Fermi saturation it may be necessary  more than $100$
oscillations. 
Due to such a slow rate of production, the issue
of back-reaction becomes important. It is very likely that simultaneous
non-perturbative production of bosons can change the picture quite 
significantly. In some sense,
hybrid model can be considered to be the safest of all the supergravity
motivated inflationary models, because the gravitino production
can stop due to back-reaction effects coming from other newly
created bosons or fermions. In other models, such as in the chaotic
models,  
where most of the particle creation  takes place in the very first few
oscillations, the issue of back-reaction hardly plays any significant role.

The layout of the paper is as follows; in section $2$, we establish the
supergravity Lagrangian and discuss the gauge fixing mechanism for the
multi-chiral field scenario; this is a mere generalization of the
$R_{\zeta}$ gauge usually used in pure gauge theories as well as an
alternative way of explaining the existence of a unitary gauge in 
supergravity theories \cite{baulieu,casalbuoni,maroto}.  In section $3$, we
discuss the quantization procedure for the gravitinos. In section $4$, we
briefly describe a general class of supersymmetric hybrid inflationary
models we use. Analytical results and discussions based on our numerical
results are presented in section $5$. The implications of these results on
the reheat temperature are discussed in section $6$.  We give a detailed
discussion on fermionic creation in hybrid model in the appendix.

\section{General supergravity Lagrangian and gauge-fixing}

In this section we describe the supergravity Lagrangian. For the 
sake of brevity, and for our purpose, we concentrate upon the
chiral-supermultiplet which contains the bosonic part, fermionic
part and the interaction terms between the fermions. 
We consider 
minimal K\"ahler potential with $K=
\phi^{i}\phi_{i}$,
where the scalar fields $\phi_{i}$ are taken to be real. The superpotential
is denoted by $W$, and the K\"ahler function is given as usual by $G =
K + \ln |W|^2$.  The choice of minimal K\"ahler potential also ensures   
$G^{i}_{j} = \delta ^{i}_{j}$. The total Lagrangian is as follows
\cite{bailin}: 
\begin{eqnarray}
\label{lag}
& &e^{-1}{\cal L}_{\rm{SUGRA}} = \frac{R}{2}+ \partial_{\mu}\phi_{i} 
\partial ^{\mu}\phi^{i} +e^{G}\left(3 - G_{i}G^{i}\right)
\nonumber \\
&-&\frac{e^{-1}}{2}\epsilon^{\mu \nu \rho \sigma}\bar \psi_{\mu}\gamma_{5}
\gamma_{\nu}D_{\rho}\psi_{\sigma}+ \frac{i}{2}\bar \chi_{i}\gamma^{\mu}
D_{\mu}\chi_{i}+\frac{i}{2}e^{G/2}\bar \psi_{\mu}\sigma^{\mu\nu}\psi_{\nu}
\nonumber \\
&+& (\frac{i}{2}G^{j}\bar \chi_{i}\pbar \phi_{j}\chi^{i} + \frac{1}{\sqrt{2}}
\bar \psi_{\mu}\pbar\phi^{i}\gamma^{\mu}\chi_{i}+ \frac{i}{\sqrt{2}}
e^{G/2}G^{i}\bar \psi_{\mu}\gamma^{\mu}\chi_{i}
\nonumber \\
&-& \frac{1}{2}e^{G/2}\left(G^{ij}+G^{i}G^{j}\right)\bar \chi_{i}\chi_{j}
+h.c. )\,,
\end{eqnarray}
where indices $i,j,...,$ represent species of the chiral multiplets
and ${\not{\!\!\partial}}
\equiv  \gamma^{\lambda}\partial_{\lambda}$. 
The derivatives $D_{\mu}= \partial_{\mu}-\frac{i}{4}\omega_{\mu mn}\sigma
^{mn}$ are covariantized with respect to gravity, where $\omega_{\mu mn}$
is the standard spin connection \cite{bailin}. 
 The gravitino field is represented by $\psi_{\mu}$, $\chi_{i}$
represents the fermions and $e$ is the determinant of the vierbein 
$e^{\mu}_{\nu}$. We have also used $\sigma^{\mu \nu}\equiv
\frac{i}{2}[\gamma ^{\mu},\gamma^{\nu}]$.  The complete Lagrangian
would also contain 
the fermion Yukawa couplings, four fermion terms, and numerous
non-renormalizable  
terms, details can be found in the literature \cite{nilles}.
Here we have neglected the torsion term in the covariantized derivative,
assuming that the gravitino production is small, so their  back
reaction can  be neglected. However, we mention that consistency of 
 supergravity theory would demand the presence of 
the torsion term in the covariantized derivative. 
These terms are suppressed by the Planck mass square and we can safely
neglect them in the linearized  gravitino equation of motion.
The Planck mass $M_P$ has
been taken to be unity and  the convention shall be retained, unless
otherwise stated.  

The above Lagrangian is invariant under local supersymmetric 
transformation laws \cite{bailin}. For spontaneous supersymmetry
breaking to occur at least one of the field vacuum expectation values
should be non-zero. In particular, for $F$-term type breaking of local
supersymmetry  one  requires
\begin{eqnarray}
\label{susyb}
\langle 0 |\delta_{\xi} \chi_{i}|0\rangle = \langle -i\pbar \phi_i \xi -
e^{G/2}G_{i}\xi \rangle \neq 0 \,,
\end{eqnarray}
where $\xi$ is the infinitesimal Grassmann-odd parameter.
The right-hand side of Eq.~(\ref{susyb}) has two explicit terms which can
break local supersymmetry. The second term is the usual $F$-term of
the scalar field whose non-vanishing vacuum expectation value induces
supersymmetry breaking.  
The first term will give a non-zero
contribution in case of a time-varying scalar background field. 
This will be the situation if we identify the background fields
$\phi_i$ with the oscillating inflaton fields. In this case local
supersymmetry is always broken during the oscillations.  

The goldstino can be identified as usual from Eq. (\ref{susyb})
\begin{eqnarray}
\label{goldstino1}
\eta = \theta_{i}\chi^{i}\,,
\end{eqnarray}
where,
\begin{eqnarray}
\label{goldstino2}
\theta_{i} = i\pbar \phi_i - e^{G/2}G_i\,,
\end{eqnarray}
and we follow the notations introduced in Ref. \cite{riotto2}.
We  are only interested in homogeneous scalar fields 
$\phi_{i}$ which are solely function of time.
The goldstino is then ``eaten'' by the gravitino in local  
supersymmetric theories. This also ensures that the gravitino 
gains the helicity $\pm 1/2$ 
components other than the $\pm 3/2$ components, and becomes massive. 
In the high energy limit it is possible to relate the helicity $\pm 1/2$
components of the gravitino to the goldstino 
 via the equivalence principle \cite{fayet,riotto2,maroto} .
In the limit when $M_{\rm{P}} \rightarrow \infty $, the helicity $\pm 1/2$
components retain the memory of the goldstino contribution and this is the
reason why the two helicities behave differently and have different
production rates  \cite{kallosh,riotto1}. 

As it can be realized by inspecting Eq.~(\ref{lag}), the gravitino 
is coupled to the fermions through the following mixing terms
\begin{equation}
\label{mix}
\frac{1}{\sqrt{2}}\bar \psi_{\mu}\left(ie^{G/2}G_i+ \pbar
\phi_{i}\right)\gamma^{\mu} \chi^{i}\,.
\end{equation} 
However, it is possible to get rid of these mixing terms by adding a
gauge-fixing term to the Lagrangian as it has already been discussed
in various places \cite{baulieu,casalbuoni,maroto}. 
In this section we mainly concentrate on the mere existence of such a
gauge-fixing term in supersymmetric theories. Regarding this, we extend
the previous  calculation made on gauge-fixing for a single-chiral
field\footnote{In Ref. \cite{maroto}, the authors have only considered
the equation of motion for the goldstino, because they were more
interested in establishing the high-energy equivalence between helicity
$\pm 1/2$ gravitinos and goldstinos.} \cite{maroto}.

Before we move onto specifying the gauge, let us introduce the projection
operators, which we need to use later on \cite{riotto2,van}.
\begin{eqnarray}
\label{proj1}
{\cal P}^{\bot}_{ij} &=& \delta
_{ij}-\frac{\theta_{i}^{\dagger}}{\theta^{\dagger} 
\theta}\theta_{j}\,, \\
{\cal P}^{\parallel}_{ij} &=& \frac{\theta^{\dagger}_i}{\theta ^{\dagger}
\theta}\theta_j\,. \label{proj2}
\end{eqnarray}
The modulus of $\theta_i$ is given by
\begin{eqnarray}
\label{goldstino3}
\theta^{ \dagger}\theta= e^{G}G^{i}G_i+ \dot \phi^{i}(t) \dot
\phi_{i}(t)= \rho +3e^G\,, 
\end{eqnarray}
where derivative with respect to time is denoted by dot,  $\rho = \dot
\phi^{i} \dot \phi_i +V$ is the energy density, 
and the scalar potential is $V=e^{G}(G^{i}G_i -3)$. 
Accordingly, the fermion $\chi_{i}$ can be split into two components 
by using the projection operators Eqs.~(\ref{proj1}) and (\ref{proj2})
\begin{eqnarray}
\chi_i &=& \chi^{\bot}_i +\frac{\theta^{\dagger}_i}{\theta^{\dagger} \theta} \eta\,,
\label{chi} \\
\chi^{\bot}_i &=& {\cal P}^{\bot}_{ij}\chi^{j}\,. \label{chip}
\end{eqnarray}
Now, with the help of Eqs.~(\ref{goldstino2}) and (\ref{chi}) the
mixing terms between the gravitino and the chiral fermions can be
recast as
\begin{equation}
\label{fmix}
-\frac{i}{\sqrt{2}}\bar \psi_{\mu}\theta_{i}\gamma^{\mu}\frac{\theta^{\dagger i}}
{\theta^{\dagger}\theta}\eta\,,
\end{equation}
where we have used
\begin{eqnarray}
\theta _i \gamma^{\mu}\chi^{i \bot} =0\,.
\end{eqnarray}
In a similar way, one can also reduce the complex conjugate part of
the mixing terms. Eq.~(\ref{fmix}) tells us about the 
direct coupling of the gravitino to the goldstino.
The equation of motion for the gravitino $\psi_{\mu}$ that follows
from the Lagrangian Eq.~(\ref{lag}) is then
\begin{eqnarray}
\label{em}
e^{-1}\epsilon^{\mu \nu \rho \sigma}\gamma_{5}\gamma_{\nu}D_{\rho}\psi_{\sigma}
+\frac{1}{2} e^{G/2}[\gamma^{\mu},\gamma^{\nu}]\psi_{\nu}\, \nonumber \\
-\frac{i}{\sqrt{2}}
\frac{\theta_{i}\gamma^{\mu}\theta^{\dagger i}}{\theta^{\dagger}\theta}\eta =0\,,
\end{eqnarray}
with an explicit term depending on the goldstino field due to the
mixing. However, we
notice that the contribution of the mixing term in Eq.~(\ref{fmix}) can
be canceled by adding to the Lagrangian the following gauge-fixing term
\begin{eqnarray}
\label{gauge1}
i\zeta \bar F \dbar F\,,
\end{eqnarray}
where $\dbar = \gamma^{\lambda}D_{\lambda}$, $\bar F =
F^{\dagger}\gamma_0$, and the gauge-fixing function is given by 
\begin{eqnarray}
\label{gauge2}
F(\psi, \eta)=\frac{\theta^{i}\gamma^{\mu}\theta^{\dagger}_i}{\theta^{\dagger}\theta}
\psi_{\mu}+\frac{1}{\sqrt{2} \zeta}\frac{1}{\dbar}\eta\,,
\end{eqnarray}
This is a mere generalization of the
$R_{\zeta}$ gauge used in pure gauge theories. Similar gauge-fixing
has been initially introduced in Ref.~\cite{bau} in the static case, where
the scalar field has been taken to its value at the minimum 
of the potential. Therefore, once we fix the gauge, the equation of
motion for the gravitino is completely free from the goldstino. 
The limit $\zeta \rightarrow 0$  corresponds to the 
unitary gauge, and it implies $\eta \rightarrow 0$. This is equivalent
to demand that no goldstino be present in the physical spectrum. 
From here onwards we will work in the unitary gauge. 

Using the gauge-fixing condition, $F(\psi,\eta)=0$, and demanding $\zeta
\rightarrow 0$, the final equation 
of motion for the gravitino, namely Eq.~(\ref{em}) in the unitary gauge 
can be written as
\begin{eqnarray}
e^{-1}\epsilon^{\mu \nu \rho \sigma}\gamma_{5}\gamma_{\nu}D_{\rho}\psi_{\sigma}
+\frac{1}{2} e^{G/2}[\gamma^{\mu},\gamma^{\nu}]\psi_{\nu}=0\,.
\end{eqnarray}
Here the mass term for the gravitino $e^{G/2}$ depends on time, due to 
presence of oscillating background scalar field $\phi_i(t)$, whose dynamics we
shall discuss later on. From now onwards we express the mass term as $m(t)$
\begin{eqnarray}
\label{mass}
m(t) = e^{G/2} \equiv e^{K/2}|W|\,,
\end{eqnarray}
where $W$ is the model dependent superpotential. Detailed  discussion will
be given in the coming sections. The above demonstration of removing the 
goldstino dependence from the gravitino equation of motion in the unitary
gauge suggets that it is possible to generalize $R_{\zeta}$ gauge
for a multi-chiral time-varying scalar background. Imposing the unitary gauge 
simplifies the gravitino field equation of motion in general and
now we will be interested in quantizing gravitino field in a cosmological 
background.

\section{Difference between helicity $1/2$ and $3/2$}

We have seen in the preceeding section that we can recognize the 
goldstino component which couples to the gravitino field in a scenario
when the dynamics of the background scalar field is also taken into
consideration. By using the gauge-fixing term in the Lagrangian, we
have shown that it is possible to cancel the gravitino-goldstino
coupling term appropriately. The final equation of motion for the gravitino
field, thus free from the goldstino, can be used to study the gravitino
production in a dynamical background dominated by an oscillating scalar
field. In this section we do not  
attempt to re-derive the equations of motion, which have already been
discussed in several papers \cite{anupam,kallosh,riotto1,riotto2},
rather we discuss few  issues and the main equations.
By studying the equation of motion for the Rarita-Schwinger field, one
notices  
that there is a free index left, which in principle can be contracted by 
at best two possible ways, say $\gamma_{\mu}$ or $D_{\mu}$, thus
giving rise to  
two constraint equations for the whole system. In presence of a
cosmological constant, the  
equations of motion for both the helicities look alike, with two simple 
constraint equations, namely $ \gamma^{\mu}\psi_{\mu}=0, D^{\mu}\psi_{\mu}
=0$. However, this is not correct in any arbitrary gravitational
background. These constraints do not hold true 
for the helicity $\pm 1/2$ case in an oscillating scalar background 
\cite{kallosh,riotto1}, even though these 
constraints continue to hold for the helicity $\pm 3/2$ case in the same
oscillating background \cite{anupam}. 
The helicity $\pm 1/2$ components gain an effective mass during the
oscillations of the inflaton, but the helicity 
$\pm 3/2$ do not seem to see the effect of curvature at all. This gain in mass
is purely due to presence of a non-trivial background curvature, which
suggests that the different helicities couple to gravity differently.

We follow the notations of Ref.~\cite{riotto1} in order to write down the
equations 
of motions for helicity $1/2$ and $3/2$ for a momentum mode $(0,0,0,|k|)$ 
projected along the $z$ direction. 
\begin{eqnarray}
\label{motion1}
\left(i\gamma^{0}\partial_{0}+i \frac{5}{2}\frac{a^{\prime}}{a}\gamma^0
-ma +k\gamma^{3}\right)\psi_{3/2}(\tau, x) =0\,, \\
\label{motion2}
\left(i\gamma^{0}\partial_{0}+i \frac{5}{2}\frac{a^{\prime}}{a}\gamma^0
-ma +kG\gamma^{3}\right)\psi_{1/2}(\tau, x) =0 \,.
\end{eqnarray}
The equations have been written in conformal time $d\tau = dt/a$, where
$a$ is the scale factor. Prime denotes derivative with respect to $\tau$,
 and $m$ is the gravitino mass Eq.~(\ref{mass}). Here $\psi_{1/2}$
and $\psi_{3/2}$ are Majorana spinors which would correspond in the flat limit
case to the $1/2$ and $3/2$ helicity states respectively (for details see
Ref.~\cite{riotto1}). 
The matrix $G$ in Eq.~(\ref{motion2}) can be  expressed in terms of $A$ and $B$
functions \cite{kallosh,riotto1}
\begin{eqnarray}
\label{matrices}
G = A+i\gamma^0 B = \frac{p-3m^2}{\rho +3m^2} +i\gamma^0 \frac{2m^{\prime}
a^{-1}}{\rho + 3m^2}\,,
\end{eqnarray}
where $\rho$ and $p$ are denoted by
\begin{eqnarray}
\label{energy}
\rho &=& \sum_{i} |\dot\phi_i|^2 + V(\phi_i) \,, \\
p &=& \sum_{i} |\dot\phi_i|^2 - V(\phi_i) \,,  \\
\end{eqnarray}
with
\begin{eqnarray}
\label{potential}
V(\phi_{i})&=& e^{K}\left(|\partial_{i}W + \phi_{i}W|^2
-3|W|^2\right)\,.
\end{eqnarray}
Here again  dot means time derivative with respect to physical
time. The potential
energy  Eq.~(\ref{potential}) is given  in terms of the K\"ahler potential
$K$ and the superpotential $W$.
In the limit when $\phi_{i} \ll 1$ (in units of the Planck mass), $A$
and $B$ can be  expressed in a simpler form
\begin{eqnarray}
\label{simp}
G= A+i\gamma^0 B = \frac{p}{\rho} +i \gamma^0 \frac{2 \dot W}{\rho}\,.
\end{eqnarray}
It is important to point out that 
in general $|G|^2=A^2 + B^2$ is time dependent, and $|G| \neq 1$. Only in the
case of a single-chiral field $|G| =1$. For multi-chiral field
scenario, in particular in the case of a 
supersymmetric hybrid model, $|G|$ departs from $1$. This makes the
quantization scheme  slightly more involved than simple  
scenarios where $|G| =1$. To proceed with the quantization we redefine $G$
in terms of the conformal time
\begin{eqnarray}
\label{red}
G = A(\tau) + i\gamma^0 B(\tau) = e^{\int\alpha d\tau}e^{2i\gamma^0
\int \mu d\tau}\,, 
\end{eqnarray}
where the coefficient in front of the overall phase represents $|G|$. 
We concentrate upon the helicity $1/2$ case, helicity $3/2$ being a simpler 
generalization of that. We expand $\psi_{1/2}$ in terms of the mode functions
\begin{eqnarray}
\label{expansion}
\psi_{1/2}&=& 
a^{-5/2}\int\frac{d^3k}{(2\pi)^{3/2}}e^{-i\vec k.\vec x}
e^{-i\gamma^{0}\int \mu d\tau} \nonumber \\
& & \times \sum _{r =1,2}(u^{r}(\tau , \vec k)a^{r}_{k}\, 
+ v^{r}(\tau,\vec k)b^{r\dagger}_{-k} )\,,
\end{eqnarray}
where, $v^{r}(\tau, \vec k)=u^{{r}^{C}}(\tau ,-\vec k)$. The spinor 
$u^{r}(\tau , \vec k)$
satisfies the following equations of motion
\begin{eqnarray}
\label{evol}
u_{\pm}^{\prime} &=& \mp im_{\rm{eff}}u_{\pm} + ik|G|u_{\mp}\,, 
\nonumber \\
m_{\rm{eff}} &=&  ma + \mu \,, \label{meff}
\end{eqnarray}
where $u^{T} =( u_{+}, u_{-})$, and for the gamma matrices we have
used the representation in which
\begin{equation}
\gamma^0=\left( \matrix{\bf{1} & 0 \cr 0 & \bf{-1}\cr }\right )\,, \\
\gamma^3=\left( \matrix{0 & \bf{1} \cr \bf{-1} & 0 \cr }\right)\,.
\end{equation}
It is possible to write down a second
order differential equation from the set of equations in Eq.~(\ref{evol}). 
\begin{eqnarray}
\label{master0}
u_{\pm}^{\prime \prime}-\frac{|G|^{\prime}}{|G|}u_\pm^{\prime}+ 
\left( k^2 |G|^2 + m_{\rm{eff}}^2 \pm i \frac{|G|^\prime}
{|G|} m_{\rm{eff}} \, \right. \nonumber \\
\left. \mp i m_{\rm{eff}}^{\prime}\right)u_{\pm} =0\,.
\end{eqnarray}
Eq.~(\ref{master0}) can be further reduced by redefining 
$ u_\pm \rightarrow e^{-\int \alpha /2 d\tau}u_\pm$
\begin{eqnarray}
\label{master}
u_{\pm}^{\prime \prime}+\left( k^2 |G|^2 + \Omega^2 \mp i \Omega ^{\prime}
\right )u_{\pm} =0\,,\nonumber \\
\Omega = m_{\rm{eff}} + i \frac{\alpha }{2}\,.
\end{eqnarray} 
or, by redefining a new time in Eq.~(\ref{master0})
$\frac{d}{d\tau} = |G|\frac{d}{dz}$,
\begin{eqnarray}
\label{master1}
\frac{d^2u_\pm}{dz^2} + \left( k^2 + \left(\frac{m_{\rm {eff}}}{|G|}\right)^2
 \mp i \frac{d}{dz}\left(\frac{m_{\rm{eff}}}{|G|}\right)\right)u_\pm =0\,,
\end{eqnarray}
where $m_{\rm{eff}}$ is defined in Eq.~(\ref{evol}), and the equation
is formally analogous to the evolution equation for a spin-1/2 fermion
in a time-varying background. It is important to
notice that all the three equations Eq.~(\ref{master0}), Eq.~(\ref{master})
and Eq.~(\ref{master1}) are equivalent, expressed in different forms.
For our numerical results we have used Eq.~(\ref{master0}), and for our
analytical treatment we can consider any of these three equations. 
In the next section it will become clear that $|G| \leq 1$ for any
general model, so that the factor $|G|^2$ in front of the momentum
squared cannot give rise to any acausal behavior.  
Nevertheless, we note that particle creation could take place
due to time variation in $|G|$. However, from Eq.~(\ref{master1}) it is
clear that this requires a non-vanishing effective mass term $m_{\rm
eff}$, even if it is merely a constant. 
Particle production takes place due to the  breaking of  the conformal
invariance. 
In our case, the violation of conformal invariance  is
not only due to presence of the gravitino mass but also due to $\mu$,
which can be related to the goldstino mass
\cite{kallosh,riotto1,riotto2}.   

In order to evaluate the occupation number we first evaluate the hamiltonian
\begin{eqnarray}
\label{ham}
H(\tau) =\int d^3k \sum_{r} E_{k}(\tau)(a_{r}^{\dagger}a_{r}-
b_{r}b_{r}^{\dagger})+F_{k}(\tau)b_{r}a_{r} \, \nonumber \\
+F^{*}_{k}(\tau)a_{r}^{\dagger}
b_{r}^{\dagger})\,,
\end{eqnarray}
in which the momentum $k$ is along the third axis, and 
\begin{eqnarray}
E_{k}&=& 2k |G| Re(u^{*}_{+}u_{-})+m_{\rm{eff}}(|u_{-}|^2-|u_{+}|^2)\,, 
\nonumber \\
F_{k} &=& 2m_{\rm{eff}}u_{+}u_{-}+k|G|(u_{+}^2 - u_{-}^2)\,, \nonumber \\
E_{k}^2 &+& |F_{k}|^2 = m_{\rm{eff}}^2 + |G|^2k^2 \,.
\end{eqnarray}
The Hamiltonian can be diagonalized with the help of a Bogolyubov
transformation.  
The new set of creation and annihilation operators which diagonalize
the Hamiltonian can be defined as
\begin{eqnarray}
\label{bug}
\hat a( k, \tau) = \alpha_k(\tau)a(k)+ \beta_k( \tau)b^{\dagger}(-k)\,,\nonumber\\
\hat b^{\dagger}( \tau)=-\beta^{*}_k(\tau)a(k)+\alpha_k^{*}(\tau)b^{\dagger}
(-k)\,,
\end{eqnarray}
where $\alpha_k$, and $\beta_k$ are the  normalized Bogolyubov coefficients,
\begin{eqnarray}
\frac{\alpha_k}{\beta} &=& \frac{E_{k}+ \omega}{F^{*}_{k}}\,, \nonumber \\
|\beta_k|^2 &=& \frac{|F_{k}|^2}{2\omega(\omega +E_{k})} =
\frac{\omega -E_k}{2\omega}\,, \nonumber \\
\omega ^2 &=& m_{\rm{eff}}^2 + |G|^2k^2\,.
\end{eqnarray}
Now, the time dependent occupation number can be written in terms of the
vacuum expectation value of the number operator 
\begin{eqnarray}
\label{num}
n(\tau) = \langle 0|N|0\rangle = \frac{1}{\pi ^2a^{3}(\tau)}\int dk
k^2|\beta_{k}|^2\,.
\end{eqnarray}
In order to solve Eq.~(\ref{master}), one needs to specify the
boundary conditions.  
Usually they are defined such that at the beginning, 
when $\tau  \rightarrow 0$, we have $|\beta_k|^2=0$, and the occupation
number $n(0) =0$,  
suggesting that there is no particle density at the initial time:
\begin{eqnarray}
\label{init}
u_{\pm}(0)&=& \sqrt{\frac{\omega \mp m_{\rm{eff}}}{\omega}} \,, \nonumber \\
u_{\pm}^{\prime}(0)&=& \mp i m_{\rm{eff}}u_{\pm}(0) +i k|G(0)|u_{\mp}(0)\,.
\end{eqnarray}
Now, we have all the tools necessary to study the production
of gravitinos in the hybrid inflationary model. Next we describe the hybrid 
model and how to estimate the occupation number.

\section{Hybrid Inflation}

The hybrid inflation potential can be derived 
from the following superpotential
\begin{eqnarray}
\label{superpot}
W= \lambda \phi ( N^2 - N_0^2)\,,
\end{eqnarray}
where $\phi$ plays the role of inflaton. During inflation the other field
$N$ is trapped in its false vacuum, $N=0$, while the $\phi$ field rolls
down towards the critical value determined by the value of $N$ at the global
minimum, $N_0$, and the coupling constant  $\lambda$ 
between $\phi$ and $N$. At this point the $N$ field rolls down from its
zero value towards the global minima and begins to oscillate around $N_0$,
while 
$\phi$ oscillates around zero. This will enable the preheating phase
of the Universe. During this period an effective potential for the
fields $\phi$ and $N$ can be derived
\begin{eqnarray}
\label{pot}
V & = & |W_{\phi}|^2 + |W_{N}|^2 \,, \nonumber \\
 & = &\lambda^2 ( N^2 - N_0^2)^2 + 4\lambda^2\phi^2N^2 \,,
\end{eqnarray}
where the subscripts denote the derivative of $W$ with respect to the fields.
The superpotential in Eq.~(\ref{superpot}) also ensures a non-vanishing
constant vacuum energy during inflation $ V(0) =\lambda ^2 N_0^4$. We should
also mention that $\lambda $ and $N_0$ act as free parameters of the
model, but they are constrained to some extent from the COBE
normalization and the tilt in the power spectrum \cite{mar1}
\begin{eqnarray}
\label{eta}
\lambda N_0 \approx 1.27 \times 10^{15}|\eta_{*}|\,GeV \,,
\end{eqnarray}
where we have taken\footnote{$\eta_{*} $ is one of the two
slow-roll parameters, usually defined in inflationary cosmology as
$\eta_{*}= (m_{P}^2/(8 \pi)) (V''/V(0))$; the ``star'' index means
that it is evaluated at least 60 e-foldings before the end of
inflation. This should not be confused with the goldstino; we have 
already defined goldstino 
with the same notation earlier.}  $|\eta_{*} | \approx 0.01$ in our
analysis, which is 
a reasonable assumption in order not to generate a sharp tilt in
the power spectrum of the density perturbation during inflation. The present 
constraint on the spectral index is  $|n-1| < 0.2$ \cite{lyth}. Hybrid
inflationary model derived from such a superpotential is known as 
F-term hybrid inflation. Slightly different version of hybrid inflation
popularly known as D-term inflation \cite{Dinflation} can be derived
from the Fayet-Illiopoulus term appearing from an anomalous   
$U(1)$ symmetry, which could provide the necessary potential energy during
inflation. Whatsoever be the cause of such a  
potential, our argument for the gravitino production is quite generic and 
will not depend upon a particular origin of the vacuum
energy. 

Due to presence of 
a single mass scale $\lambda N_0$ in the model, which is related to the
supersymmetric breaking scale during the inflationary era,  we will have a 
single frequency of oscillations during the 
preheating phase, and  effectively a single scalar
field oscillating. In our case this can be taken as $N$, related to the other
field $\phi$ by
\begin{eqnarray}
\label{relation}
\phi = \frac{N_0 - N(t)}{\sqrt{2}}\,.
\end{eqnarray}
By solving the equation of motion for $N(t)$  \cite{mar,mar1} we get
the approximate solution when the field is around the bottom of the potential
\begin{eqnarray}
\label{sol}
\frac{N(t)}{N_0} \approx 1 + \Sigma(t)\cos(\nu_{\phi}t)\,,
\end{eqnarray}
where $\nu_{\phi} = 2\lambda N_0$  provides the natural frequency of the
oscillations, and $\Sigma(t)$ is the amplitude of the oscillations 
decreasing in time as $\sim 1/t$ \footnote{The actual decrease in amplitude over an
oscillation period depends on the ratio $H/\nu_\phi$, $H$ being the
Hubble parameter.}.
The above expression is valid when $|N(t)/N_0 -1|\leq 1/3$, with $\Sigma (0) \sim 
1/3$. 
What makes the hybrid scenario interesting is that $N(t)$ never
vanishes and that causes the fermionic creation to be completely different
than the usual chaotic type inflationary potentials.

Now, with the present knowledge we can evaluate $|G|$ with the
help of Eq.~(\ref{simp})
\begin{eqnarray}
\label{modg}
|G|^2 & =& \frac{(\rho -2V)^2}{\rho^2} + \frac{4|\dot W|^2}{\rho^2}\,,
 \\
&=& 1 - 4 \frac {| W_\phi \dot N - W_N \dot \phi |^2}{\rho^2} \,,
\label{g2} \\ 
& =& 1 - \frac{4\lambda ^2 N_0^2}{\rho^2}|\dot N|^2 (N- N_0)^2 \,,
\end{eqnarray}      
where the subscript in $W_\phi$, $W_N$ means derivative with respect to
the corresponding field, and dot denotes as usual physical time derivative. 
The last equation has been written using 
Eq.~(\ref{relation}), knowing that $ |\dot \phi|^2 + |\dot N|^2 =
3/2 |\dot N|^2$. Here it is important to notice that the departure from
$1$ is quite obvious, even though there is effectively a single scalar
field oscillating. In fact, we can easily generalize  Eq.~(\ref{g2}) to any
arbitrary number of scalar  fields $\phi_i$ contributing to the energy
density. Using Eqs.~(\ref{matrices}-\ref{potential}), without assuming
$\phi_i \ll 1$, we get
\begin{equation}
\label{proof}
|G|^2 = 1 - \frac{4 e^K}{(\rho+3 m^2)^2} \sum_{i<j} | {\cal D}_i W \dot \phi_j -
{\cal D}_j W \dot \phi_i |^2\,,
\end{equation}
where ${\cal D}_i W= W_i + K_i W$, and the subscript $i$ denotes
derivative with respect to the homogeneous scalar field $\phi_i$. 
This shows explicitly that $|G| \leq 1$.  

\section{Analytical estimation of occupation number}

\subsection{Helicity $1/2$}

It is possible to analytically estimate the occupation number, and the
number density $n(t)$ for both the helicities.
Assuming that $|\beta_{k}|^2 \approx 1$, for a given momentum $k$, our task 
reduces to estimate the cut-off momentum $k$. 
For this we need to
solve Eq.~(\ref{master}), which will then lead us to study the
occupation number for 
the helicity $1/2$ gravitinos in hybrid model. In the case of 
helicity $1/2$ states, the contribution of $|G|^2$ can be important, and we 
need to evaluate it before we could estimate the abundance of 
helicity $1/2$ states. Later on we will discuss the
abundance of helicity $3/2$ states, where $|G|^2 =1$.
To carry out our calculation we need to know the dominant contribution 
to $\Omega$ appearing in Eq.~(\ref{master}),
\begin{eqnarray}
\label{known}
\Omega &=& ma + \mu + i \frac{\alpha }{2}  \,,
\end{eqnarray}
where
\begin{eqnarray}
\mu &=& - \frac{A^\prime}{2B} + \frac{A}{2B}\frac{|G|^\prime}{|G|} \,,
\\
\alpha &=& \frac{|G|^\prime}{|G|}\,.
\end{eqnarray}
With the help of Eq.~(\ref{sol}) and Eq.~(\ref{modg}) it is possible to
evaluate all the 
terms in Eq.~(\ref{known}). Here we simply quote the final results
in terms of the physical time:
\begin{eqnarray}
\label{mass1}
m(t)&=& -\frac{\nu_{\phi}^3\Sigma^2(t)}{4\sqrt{2}\lambda ^2 M_{\rm p}^2}
\left(1 + \cos(2\nu_{\phi}t)\right)
+ {\cal O}({1}/{t^3})\,, \\
|G|^2&=& 1- \frac{1}{9}\sin^2(2\nu_{\phi}t) +{\cal O}({1}/{t})\,,
\label{g2app} \\
\frac{|\dot G|}{|G|}&=& - \frac{\nu_{\phi}}{9}\sin(4\nu_{\phi}t) 
\left(1 +\frac{\sin^2(2\nu_{\phi} t)}{9}\right ) + {\cal O}({1}/{t})\,, \\
\label{dom}
\frac{\dot A}{2B}&=& \frac{3\nu_{\phi}}{2\sqrt{2}}
\left(1 +\frac{5\Sigma (t)}{4} \cos (\nu_{\phi}t)\right) +
{\cal O}({1}/{t^2}) \,, \\
\label{subdom}
\frac{A}{2B}\frac{|\dot G|}{|G|}&=& \frac{\nu_{\phi}}{6\sqrt{2}}
\cos^2(2\nu_{\phi}t)\left(1+\frac{\sin^2(2\nu_{\phi}t)}{9}\right)
+{\cal O}({1}/{t}) \,.
\end{eqnarray}
It is apparent that the only term which dominates $\Omega (t)$ in
Eq.~(\ref{known}) is due to Eq.~(\ref{dom}). The mass term $m(t)$ is subdominant due to
the Planck suppression ( here we have explicitly written the Planck mass). 
The amplitude of the oscillations from Eq.~(\ref{subdom}) is one order of magnitude 
smaller compared to that in Eq.~(\ref{dom}). As a first approximation we can 
consider $\dot A/(2B)$ to be the  effective mass term for the helicity $1/2$ states. 

\begin{figure}[t]
\epsfxsize=8cm
\epsfysize=8cm
\hfil \epsfbox{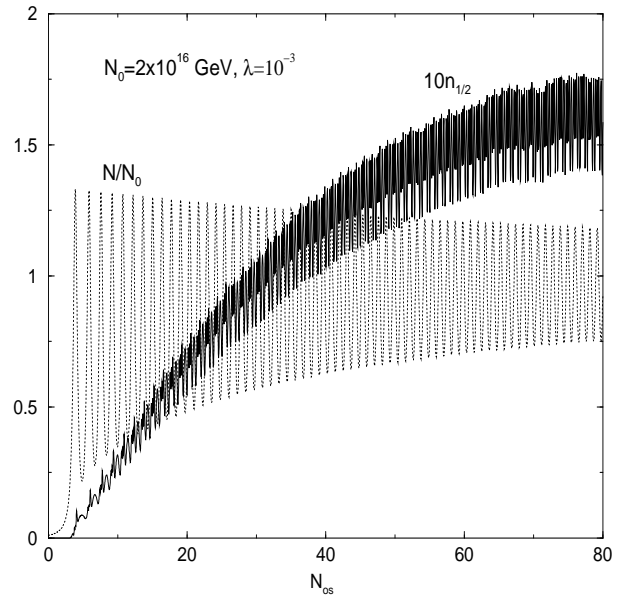} \hfil
\caption{\label{mar4} \small{The evolution of the classical field $N(t)$, and the
number density	of helicity 1/2 gravitinos (scaled by a factor of 10)
has been depicted for the   
choice of parameters $\lambda=10^{-3}$ and $N_0=2 \times 10^{16}$
GeV. The time scale is given in terms of the approximate number of
oscillations of the fields in physical time, $N_{\rm os}=\nu_\phi t/2\pi$. 
The number density is given in units of $\nu_\phi^3$.}}
\end{figure}

\begin{figure}[t]
\epsfxsize=8cm
\epsfysize=8cm
\hfil \epsfbox{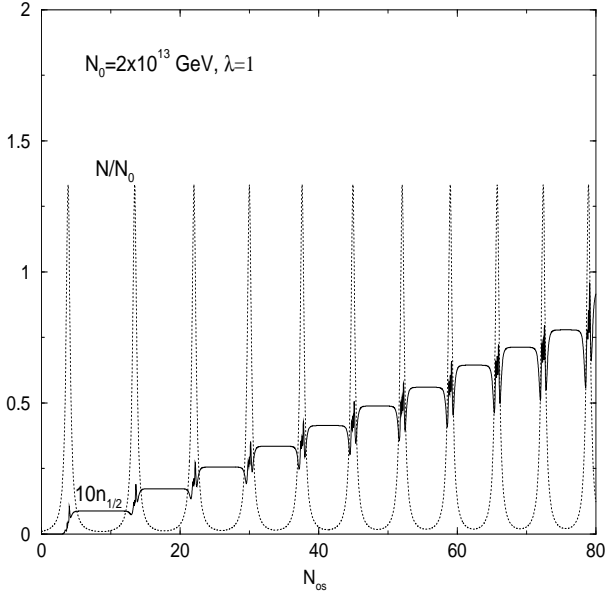} \hfil
\caption{\label{mar5} \small{The evolution of the classical field $N(t)$, and the
number density	of helicity 1/2 gravitinos (scaled by a factor of 10)
has been depicted for the choice of parameters $\lambda=1$ and
$N_0=2 \times 10^{13}$ GeV. The number density is given in units of
$\nu_\phi^3$. We  note that the rate of production depends mainly on 
$\nu_{\phi}$, and as a result the production rate is similar to that of
Fig.~(\ref{mar4}).}}
\end{figure}

A general discussion of production of massive fermions in an
oscillatory background is given in the Appendix. We will use the
results quoted there in order to estimate the occupation number for 
the helicity $1/2$ gravitinos. By taking $\Omega (t)\approx -\dot A/(2B)$, and 
$|G|^2 \approx 1$, we notice that Eq.~(\ref{master})
mimics Eq.~(\ref{sec}) in the appendix for a time-varying mass which is denoted
by $\Omega(t)$ in our present situation.  First of all we notice that the effective 
mass term $\dot A/(2B)$ never vanishes at any point of the scalar field
oscillations. With $\Sigma(t) \sim 1/3$, the effective mass is always 
positive. This situation is typical of any hybrid model. Comparing Eq.~(\ref{dom})
with Eq.~(\ref{mass2}) in the Appendix, we have 
$\nu_\phi=m_\phi$, $m_X =3\nu_{\rm{\phi}}/(2\sqrt{2})$ and 
$g\phi(0) = 5\nu_{\rm{\phi}}/(8\sqrt{2})$. It is evident that $m_X > g\phi(0)$, and as
a result $m(t)$ is always positive, quite similar to the case of bosonic
production, where the mass term appears always as squared.
However, when studying fermionic creation for a chaotic type potential 
there is a possibility 
of having $m(t)$ vanishing at some point during the oscillations
\cite{peloso}.  In this case fermion production takes place in  
the first few oscillations, since the
adiabatic condition is violated maximally when the inflaton field passes 
through the point where the effective mass vanishes, and it is
possible to create very heavy fermions in the process. 
In hybrid models and for the helicity 1/2 gravitinos, because  the
effective mass remains always positive, 
the adiabatic condition is broken when the effective mass reaches its
$minimal$  (non zero) value, and the mass of the fermions created
never exceeds  the mass of inflaton. In this case it is also important
to note 
that the amplitude of the oscillations decays slowly and it takes roughly
$20-30$ oscillations to make any significant change in the initial amplitude.
As a result the production of fermions takes place at each and every
oscillation,  
and since the degree of violation of the adiabaticity is much weaker
compared to  the former, the production process takes a longer time to
saturate the Fermi band.  
This is quite evident from our numerical results, see
Fig.~(\ref{mar4})  
and Fig.~(\ref{mar5}). The  choice of model parameters in
Fig.~(\ref{mar4}) leads  
to a higher inflationary scale compared to that of Fig.~(\ref{mar5}). 
However, the rate of 
production is exactly the same in both the cases. The reason is that 
the occupation number essentially depends on $\nu_{\phi}$, which is
exactly the  
same in both the models we have considered. This can be vividly seen
by comparing the number of peaks in the occupation number.

In Fig.~\ref{mar5} it is easy to see the particle creation taking
place in bursts  
and in regular intervals. The peaks remind us the violation of the
adiabatic condition.  
This is a perfect example of particle creation in a broad resonance
regime. In the  
narrow resonance, particles are always created throughout the evolution of the 
scalar field \cite{linde}, they are not produced in bursts like in Fig.~\ref{mar5}.
It is visible 
that the particle number remains constant for some time and then 
there is a sudden jump in the
occupation number. 
Physically one can 
understand the situation from the comparison with an interacting
quantum field theory. Usually in field theory, 
at extreme past/future  we consider the plane wave solution of the free
equation of motion. However, in between
extreme past and  future the interaction is switched on. Here also as
we can see 
that the occupation number remains constant for a while, which represents the 
ingoing wave and 
then the interaction switches on, which is depicted by a jump in the 
occupation number. The 
outgoing wave in this picture is the newly filled occupation number. In 
fact the Bugolyubov
coefficients can be calculated by estimating the reflection and the 
transmission coefficients 
of the plane wave passing through a barrier. Most of the important
information  
concerning  the production of helicity $1/2$ gravitinos can be
extracted  from  
Fig.~\ref{mar5}. The production of helicity $1/2$ gravitinos builds up
in each and  every oscillation. The production does not saturate the
Fermi level in the first few oscillations, 
but it takes several oscillations to reach the Fermi level. This behaviour is 
in stark contrast 
to the fermionic creation in a quadratic potential \cite{peloso}, where the 
Fermi level is reached 
in a few oscillations. It is also noticeable that there is no stochastic 
behaviour in the occupation number, and it is always increasing until
it reaches the Fermi level. The main reason is that the effect of
expansion is felt  very slowly in 
hybrid models, when compared to the typical period of oscillation,
that is, $H_0/\nu_\phi \approx N_0/M_P \ll 1$, where $H_0$ is the
value of the Hubble constant at the end of inflation. 
Therefore,  the amplitude decreases very slowly in most of the cases, 
almost adiabatically.
There is also a slight change in frequency, as it can be noticed from
the plot (for details see 
Ref.~\cite{mar1}). However this small change in frequency is not going to
affect our analytical estimation.

\begin{figure}[t]
\epsfxsize=8cm
\epsfysize=8cm
\hfil \epsfbox{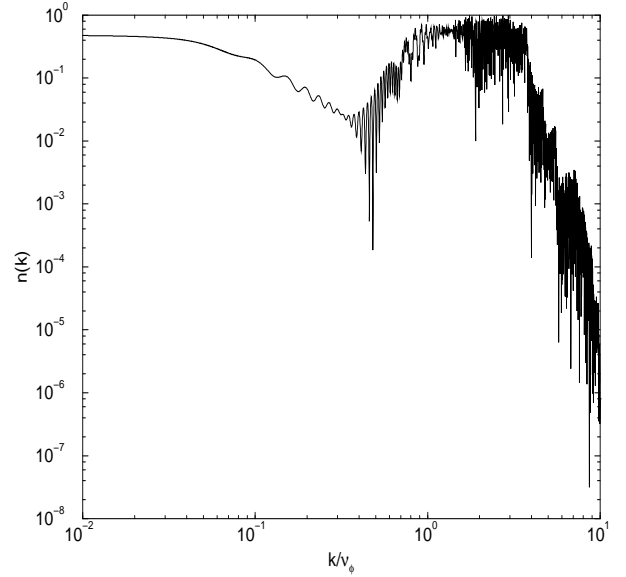} \hfil
\caption{\label{mar1} {The spectrum of helicity  $1/2$, when   
the number of oscillations is $N_{\rm os}=60$, for the model parameters
$\lambda=10^{-3}$ and $N_0=2\times 10^{16}$ GeV. This choice of model 
parameters leads to $\eta_{*} \sim 0.01$. It is evident that the
Fermi level is saturated for a cut-off momentum $k \sim \nu_{\phi}
\sim 2 \lambda N_0$. }}
\end{figure}

\begin{figure}[t]
\epsfxsize=8cm
\epsfysize=8cm
\hfil \epsfbox{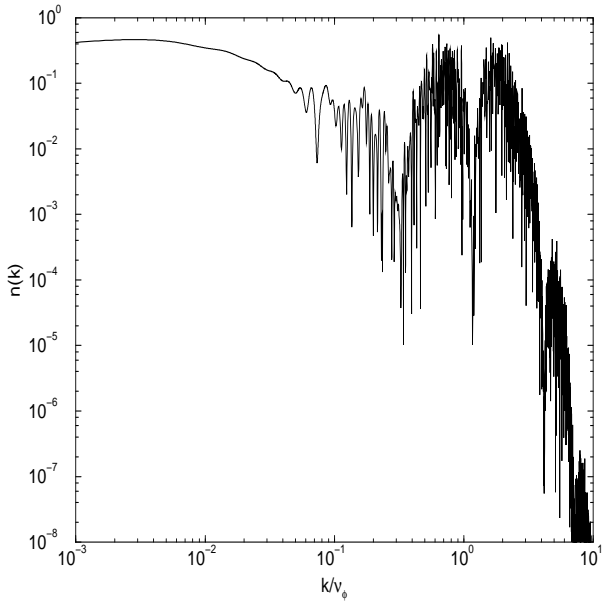} \hfil
\caption{\label{mar2} {The spectrum of helicity $1/2$, when  
the number of oscillations is $N_{\rm os}=60$, for the model parameters
$\lambda =1.0$ and $N_0 = 2 \times 10^{13}$ GeV. This choice of model
parameters 
leads to $\eta_{*} \sim 0.01$. It is evident that the
Fermi level is saturated for a cut-off momentum $k \sim \nu_\phi
\sim 2 \lambda N_0$. Important point to notice that this choice of
model parameters leads to low inflationary scale and this gives a 
different spectrum than noted earlier in Fig.~(\ref{mar1}).}}
\end{figure}

We have plotted the spectrum for the helicity 
$1/2$ gravitinos in Figs. (\ref{mar1}) and (\ref{mar2}), for two
different set of model parameters. In both the 
cases Fermi-level is saturated for $k_{\rm{max}} \approx \nu_{\rm{\phi}}$. 
This confirms our
analytical study in the preheating section, see Eq.~(\ref{estim1}),
which reduces to  numerically observed $k_{\rm {max}}$ for $m_X \sim
m_{\phi}$. 
The number density of helicity $1/2$ can be obtained  from  Eq.~(\ref{num})
\begin{eqnarray}
\label{esti1}
n_{1/2} \approx \frac{1}{4\pi^3}\int d^3k n(k) \simeq \frac{k^3
_{\rm{max}}}{3\pi^2} \approx \frac{\nu_{\rm{\phi}}^3}{3\pi^2}\,.
\end{eqnarray}
Here $k_{\rm{max}}$ has been taken to be a comoving momentum.
It is evident from Fig.~(\ref{mar4}) and Fig.~(\ref{mar5}) that
the occupation number grows gradually and saturates after many oscillations
depending on the choice of $\lambda$ and $N_0$.

\begin{figure}[t]
\epsfxsize=8cm
\epsfysize=8cm
\hfil \epsfbox{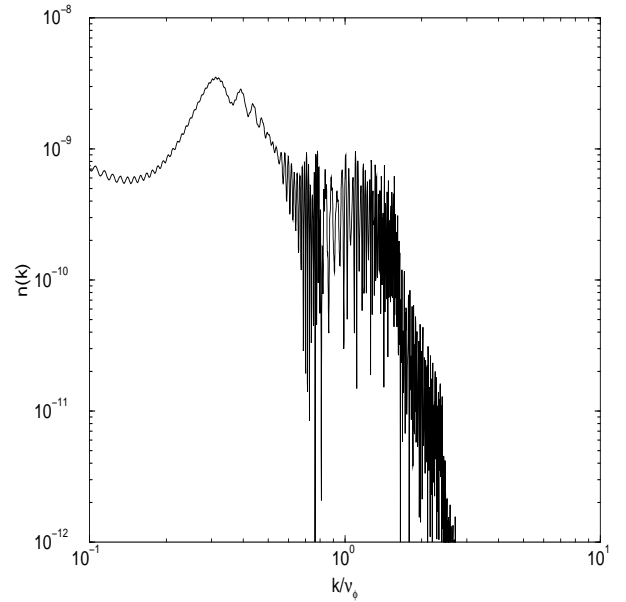} \hfil
\caption{\label{mar3} \small{Spectrum of helicity $3/2$, when 
the number of oscillations is $N_{\rm os}=60$, for the model parameters
$\lambda =10^{-3}$ and $N_0 =2\times 10^{16}$ GeV. This choice 
of model parameters
leads to $\eta_{*} \sim 0.01$. It is evident that the
Fermi level is not saturated in this case. Even though the production is 
small compared to helicity $1/2$, it is certainly not negligible. Important 
thing to notice that the spectrum preserves its shape and peaks around
$0.3 \nu_\phi$.}} 
\end{figure}
\subsection{Helicity $3/2$}

We can follow similar arguments to evaluate the occupation number
for helicity $3/2$ gravitinos. Notice that Eq.~(\ref{motion1})
for helicity $3/2$ reduces exactly to Eq.~(\ref{master}) with
$|G| =1$ and $\Omega = m_{\rm{eff}} = m a$,
\cite{anupam,kallosh,riotto1,riotto2}. Hence, the single time-varying mass
scale appearing in the problem is now given by Eq.~(\ref{mass1}). Comparing
Eq.~(\ref{mass1}) with   
Eq.~(\ref{mass2}), it becomes clear that the bare mass $m_{X}$
itself is time-varying
for the helicity $3/2$ case, but as we have noted before, the
amplitude of the oscillations is almost constant, especially in the
hybrid model we are interested in. 
The parameters $m_X $
and $g\phi(0)$ are equal to each other, 
and both are Planck mass suppressed. But this does not mean that
we cannot excite helicity 3/2 gravitinos.
The main point is that the effective mass
 always vanishes at some point during each and every oscillation.
 As a matter of fact we can
excite them  precisely due to this. What
matters is the violation of adiabaticity, and that takes place 
precisely at those points where $m(t)$ vanishes. Production of
helicity $3/2$ mimics the first scenario we have discussed in
the appendix, but now  $m_X \ll m_{\phi}=
2\nu_{\phi}$. However, since the effective mass $m(t)$ vanishes 
in each and every oscillation, so the gravitino production takes place
in each and every oscillation.  The spectrum of helicity $3/2$ gravitinos
is plotted in Fig.~(\ref{mar3}), for the model parameters given in
Fig.~(\ref{mar1}).  
The spectrum preserves the essential features obtained before, but most
importantly Fermi level never gets saturated and the production is extremely 
subdominant compared to helicity $1/2$ case. Nevertheless, the helicity $3/2$
gravitino abundance may also pose a 
strong bound on the model parameters and the reheating temperature, and
thus it is necessary  to study them as well \cite{anupam}. Our task is to
estimate $k_{\rm{\max}}$, and with the help of Eq.~(\ref{con1}) we get
\begin{eqnarray}
\label{trans}
k^3_{\rm{max},3/2}\approx \frac{\nu_{\phi}^5 \Sigma^2(t)}{2\sqrt{2}\lambda^2
M_{\rm p}^2}\,.
\end{eqnarray}
By taking $M_{\rm p} \approx 10^{18}$ GeV, and $\nu_{\phi} \approx 10^{14}$
GeV, we get 
$k_{\rm{max}} \approx 0.1 \nu_{\phi}$, which matches very well with
our numerical result. In  Fig.~(\ref{mar3}), the spectrum peaks around 
$0.3 \nu_{\rm{\phi}}$. It is straightforward to estimate the
occupation number for the helicity $3/2$ gravitinos
\begin{eqnarray}
\label{esti2}
n_{3/2} \approx \frac{|\beta|^2\nu_{\phi}^5\Sigma^2(t)}{6\sqrt{2}\pi ^2 \lambda ^2 
M_{\rm p}^2}\,.
\end{eqnarray}
Here we have assumed $\beta_k$ in Eq.(\ref{num}) to be a constant, $0 \leq
|\beta|^2  
\leq 1$, for the cut-off momentum $k_{\rm{max}}$. The actual value of
$\beta $ is 
difficult to estimate analytically, and we do not attempt to analyse it here.
We can also  estimate the abundance ratio for the two helicities
\begin{eqnarray}
\label{rati}
\frac{n_{1/2}}{n_{3/2}} = \frac{2\sqrt{2}\lambda ^2 M_{\rm p}^2}
{|\beta|^2\nu_{\phi}^2\Sigma^2(t)}\,.
\end{eqnarray}
We should mention that in Eqs.~(\ref{trans}-\ref{rati}), 
$\Sigma(t)$
can be taken to be $1/3$, since the amplitude of the oscillations 
remains unchanged for many oscillations. For the numerical values we
have considered, we would get the production of helicity $1/2$ gravitinos
to be  roughly $4$ orders of magnitude larger than that of the helicity
$3/2$ states, if we naively 
assumed that helicity $3/2$ saturated the Fermi-level ($|\beta|^2= 1$,
which is an over estimated production). It is evident from the
numerical  
example, that the production is suppressed  by at least
$8$ orders of magnitude compared to that of helicity $1/2$, see
Fig.~(\ref{mar1}) and Fig.~(\ref{mar3}).

\section{Implications on $T_{\small{\rm{reh}}}$}

Through parametric resonance, other particles can also be created, and 
especially if their couplings are not suppressed
by the Planck mass, they are perhaps produced more abundantly than gravitinos.
Assuming that the end of reheating gives rise to a thermal
bath with a final temperature $T_{\rm{reh}}$, it is possible to estimate
the ratio  $n/s$, where $s$ is the entropy density, and $n$ represents
the number density of gravitinos after the end of reheating. Both $n$
and $s$ scales like $a^{-3}$, and  we can also assume that the
background scalar field density behaves on average like matter, $\rho_{\phi}
\propto a^{-3}$,  during the oscillatory period.  
It is possible then to estimate the final ratio by noticing that
$\rho_i \approx \lambda^2 N_0^4 $ and $n(t_{\rm{i}}) \approx \lambda^3N_0^3$,
where the subscript denotes the initial values
\begin{eqnarray}
\label{ratio}
\frac{n}{s} = \frac{n(t_{\rm i})}{\rho_i}T_{\rm{reh}}\approx
 \frac{\lambda}{N_0} 
 T_{\rm{reh}} \,.
\end{eqnarray}
Here the left hand side represents the final abundance of gravitinos during nucleosynthesis.
Gravitinos are weakly coupled to gauge bosons and its gaugino partners and
their lifetime, $\tau_{\rm{decay}} \approx M_{\rm p}^2/m_{3/2}^2$, is
very long. For a TeV mass gravitino it could be around $10^4-10^5$
seconds, and it would pose a genuine threat to nucleosynthesis. However, 
this statement is strictly correct only for the
helicity $3/2$ component, since they can decay to gauge bosons
and its gaugino partner through a dimension $5$ operator. At 
high energies the interaction channels are governed by $3/2$
component rather than $1/2$ component gravitinos. In particular, 
helicity $3/2$ may be produced with a mass close to a TeV range, so they
decay very late and they are the ones which survive till late to cause 
problems for nucleosynthesis. However, the same can not be said 
with confident about the helicity $1/2$ components. As 
we have seen, the violation of conformal invariance is not the same 
for both the helicities, and the helicity $1/2$ gains an effective mass
which is 
of the order of $\nu_{\phi} \gg m_{3/2}$. Essentially, helicity $1/2$
gravitinos are in an oscillatory scalar background with a frequency
similar to their effective mass. There
is no good reason to believe that the decay rate of helicity $1/2$ to gauge
bosons and to gauge fermions would mimic the decay rate similar
to that in a flat background. So far, a detailed study is lacking in this
area, but there is sufficient hint that the decay rate of helicity
$1/2$ is much smaller than the Hubble parameter. We do not repeat the 
argument, rather we refer the reader to  Ref.~\cite{riotto1}. The detailed
calculation  of the decay rate seems to be quite involved and we leave that
for  our future investigation. The important point to realize is that once the
Universe reheats and thermalizes, the effective mass of the helicity $1/2$
gravitinos becomes similar to that of the helicity $3/2$,  and as a result
the decay rate would essentially be given by the usual decay rate in a flat
background limit. 
Whatsoever be the detailed analysis, we must mention that while
deriving Eq.~(\ref{ratio}), we have implicitly assumed that the initial
abundance of gravitinos produced remain frozen until thermalization.
Since the decay rate is smaller than the Hubble rate, the gravitinos
produced during preheating will be able to survive until the end of reheating.
The final abundance will mainly depend on 
the model parameters such as $\lambda $ and $N_0$, see Eq.~(\ref{ratio}).
Therefore, the constraint on the
reheat temperature derived from Eq. (\ref{ratio}) is
clearly model dependent. 

We also mention that we have not included the effect of 
back-reaction coming from newly created bosons and fermions. 
In hybrid models, the production of the quanta
associated with $N$ and $\phi$ is very efficient and can take place
just in the first few oscillations \cite{mar1}, much before the gravitinos have 
saturated  the Fermi level. 
Therefore, back-reaction
effects due to these quanta will quickly change the frequency and also the
amplitude of the oscillating fields.
We strongly suspect that especially in the hybrid scenario gravitino production
will be affected due to such considerations and hence our current
estimation of gravitino abundance will not hold anymore. However, even though
the non-thermal production of gravitinos may stop after a while, there is a
possibility to create them through scattering processes. This is beyond the 
scope of the present discussion and we shall hope to come back to these issues
elsewhere \cite{ma}.


\section{Discussion and conclusions}

We have carried out the calculation for the gravitino production in  
multi-chiral 
field scenario, in particular in the context of hybrid model. As we have 
shown, it is possible to add a gauge-fixing term to the supergravity
Lagrangian  
to get rid of the mixing between the goldstino and the gravitino field.
Our method is a generalization of $R_{\zeta}$ gauge studied in various
contexts. We choose to work in a unitary gauge, in which the goldstino is
completely removed from the physical spectrum. 
Our study emphasizes
major points in the non-perturbative production mechanism of
gravitinos, 
analytically and numerically for the multi-chiral field models. We also
give detailed analysis  
of the fermionic creation in general. We have observed that the fermionic creation
in hybrid model is quite  
different from other chaotic inflationary models. The effective fermionic
mass in hybrid models  never vanishes, and as a
result the particle production does not take place in first few
oscillations, 
rather it builds up gradually. This makes it more interesting as far
as the gravitino production is concerned. 
If we really want nucleosynthesis to be preserved in the context of 
supergravity inflationary models, we believe models based on 
hybrid inflation with low scales are probably going to be the only saviour.
The reason is very simple; in other models there is no way we can 
argue the back-reaction due to the creation of other particles would 
stop creating gravitinos, but in hybrid models there is a scope where 
the back-reaction due to non-perturbative creation of bosons could
affect the coherent oscillations of the inflaton and halt the particle production
completely. This gives us a new hope to understand the abundance of 
gravitinos during nucleosynthesis and we leave these important issues 
to be investigated in near future.

\section*{Acknowledgements}

A.M. is supported by Inlaks foundation and the authors are thankful to 
Lev Kofman, David Lyth, Antonio Lopez Maroto, and Subir Sarkar  
for many useful suggestions.

\appendix
\section*{Estimation of $k_{\rm{max}}$}

In this section we briefly discuss non-perturbative production
of massive fermions. This discussion is general and self sufficient.
We begin with the second order Dirac equation represented in terms of
the two mode functions. It reads in conformal time
\begin{eqnarray}
\label{sec}
u_{\pm}^{\prime \prime} +\left( k^2 + (ma)^2 
\pm i(ma)^{\prime}\right)u_{\pm} =0\,,
\end{eqnarray}
where $u_{\pm}$ are the mode functions, $k$ is the momentum, 
prime denotes the derivative with respect to the conformal time, $a$
denotes the scale factor, and $m$ is an effective time-dependent
mass. This mass can be  written in terms of the physical time as
\begin{eqnarray}
\label{mass2}
m(t) &=& m_{X} + g\phi(t)\,, \\
\phi(t)&=& \Phi(t) cos(m_{\phi}t)\,, \label{phit}
\end{eqnarray}
where $m_{X}$ is the mass of the fermion we are interested in,
$\phi(t)$ is the scalar field oscillating 
with some initial amplitude $\Phi(0)$, and $m_\phi$ is its mass. The
Yukawa coupling between the 
fermion and the scalar is determined by $g$.
Inspecting the mode equation Eq.~(\ref{sec}), it is obvious that it
mimics harmonic oscillator with a time-varying imaginary frequency
\begin{eqnarray}
\omega^2 = k^2 + (ma)^2 \pm i (ma)^{\prime}\,.
\end{eqnarray}
Particle creation occurs due to  
non-adiabatic evolution of the total frequency $\omega(t)$, that is,
whenever 
\begin{eqnarray}
\label{adia}
\frac{d\omega(t)}{dt} \geq \omega^2\,,
\end{eqnarray}
where we have expressed the total frequency in physical time. 

We shall explore here two possible scenarios for fermionic 
production, depending on whether or not the effective mass for the
fermion vanishes at some point of the scalar field oscillations. 
To start our discussion, we consider first the situation when the amplitude
of the scalar field initially satisfies $|\Phi(0)| > m_{X}/g$, so that
the effective mass goes through zero at some point during the oscillations. 
Since the particle production takes place within a time $\Delta t \ll H^{-1}$,
we can safely neglect the effect of expansion during the production of
particles, 
especially in a broad resonance regime\footnote{In the hybrid model scenario
presented in this paper, it is evident from
Fig.(\ref{mar5}) that 
the gravitino production is taking place in bursts within a short interval
of time.  
However, there is another possibility, a narrow resonance regime, where the 
particle production takes place through out the evolution of the
scalar field  \cite{linde,peloso}.}.
However, we have to keep in mind that the amplitude is gradually
decreasing due to the effect of expansion. 
Once the amplitude drops below a critical value, the 
production of the fermions will soon stop and their number density will be
frozen in time \cite{peloso}. 
The critical amplitude is given approximately by equating
Eq.(\ref{mass2}) to zero
\begin{eqnarray}
\label{amp}
\Phi_{*} \simeq \frac{m_{X}}{g}\,.
\end{eqnarray}
Below this value, the effective mass remains always positive. It is
also important to mention that the production enhances near the 
regime when the effective mass vanishes, or, in other words, when there
is a maximum violation of adiabaticity condition. 
With the above information it can be possible to estimate the typical momentum
$k$ required to violate the adiabatic condition when the amplitude of
the field is close to $\Phi_*$. This will occur  when $ \cos(m_{\phi}
t_*) \sim -m_{X}/(g\Phi_*)$, or $ t_* \sim \pi/m_{\phi}$.
Our condition  Eq.~(\ref{adia}) implies
\begin{eqnarray}
\label{con}
2g^2\dot \phi (t)^2\left(m_{X} + g\phi(t)\right)^2 + \frac{1}{2}
g^2\ddot\phi(t)^2
\geq \, \nonumber \\
\left((k^2 +(m_X +g\phi(t))^2)^2 +g^2 \dot \phi(t)^2\right)^{3/2}\,.
\end{eqnarray}
Since, $\dot \phi (t_*) = 0$, and $\ddot \phi(t_*) 
\approx m_{\phi}^2 m_{X}/g$, we can estimate the left-hand side of 
Eq.~(\ref{con}) around $\phi(t_*)$, and the final condition translates
to a simpler form,
\begin{eqnarray}
\label{con1}
\frac{1}{2}m_{X}m_{\phi}^2 \geq k^3_{\rm{max}} \,.
\end{eqnarray}
We have assumed $ \omega \approx k_{\rm{max}}$ in the final 
derivation. This result
confirms similar result already obtained in Ref.~\cite{peloso}.
We can express Eq.~(\ref{con1}) in terms of  an effective $q$ parameter
\begin{eqnarray}
\label{q}
k_{\rm {max}} \sim \left ( \frac{m_{\phi}^4}{m_{X}}q_*\right)^{1/3}\,, 
\end{eqnarray}
where $q(t)= g^2 \Phi^2(t)/m_{\phi}^2$. In the above equation the value of $q$
is evaluated at $\Phi \sim \Phi_{*}$. The $q$ dependence in $k_{
\rm{max}}$ is quite different from the bosonic production. This  
is mainly due to the presence of the imaginary part of the frequency,
which has a significant contribution to the violation of the adiabatic
condition. 
At this point one may be able to estimate the maximum mass $m_{X}$ allowed
by the violation of the adiabatic condition.
With the help of the $q$ parameter, it is possible to re-express $m(t)$ as
\begin{eqnarray}
m(t)= m_{X} + \frac{\sqrt{q(0)}}{t}\cos(m_{\phi}t)\,,
\end{eqnarray}
where we have written explicitly the time-dependence of the
amplitude. 
Hence, the maximum mass is achieved when $t_* = \pi/m_{\phi}$, and this gives
\begin{eqnarray}
m_{X} \leq \frac{m_{\phi}}{\pi} \sqrt{q(0)}\,.
\end{eqnarray}
For reasonable values of the coupling constant $g$, it is possible to
achieve very high values of the fermion mass $m_X \gg m_{\phi}$. This
suggests that such production of supermassive fermions is indeed
non-thermal and non-perturbative in nature. It is also important to notice
that for values of $q$ which are of the order of tens, the maximum fermionic
mass obtained is of the order of the mass of the oscillating field.

The above analysis suggests that creating super-massive fermions is
possible because the effective mass $m(t)$ vanishes around $\phi_{*}$, and
this is where the adiabatic condition is violated maximally. This happens
quite naturally in quadratic inflationary potentials because the initial
amplitude of the oscillations for the inflaton is large enough to pass 
through the point where the effective mass $m(t)$ vanishes
\cite{peloso}. A similar situation arises in the case of the helicity
$3/2$ states for the gravitino, and  
we find that its mass term Eq. (\ref{mass1}) vanishes 
in each and every oscillation of the $N$ field, defined earlier. 
This means that the adiabatic condition is violated maximally at those
points. However,  
in the hybrid case the amplitude of the oscillations and the mass of
the helicity $3/2$ states are exactly the same $m_{X}=g\Phi(t)$. This
suggests that we can not create 
very massive $3/2$ states, although we can create them with mass $m_X
\ll m_{\phi}$. Therefore, the hybrid situation is slightly different
from the chaotic inflationary scenario with quadratic potential.

Next, we study a scenario where the effective mass of the fermion
never vanishes at any point, that is  $ g\Phi(0) < m_X$.
This is the situation which arises in the hybrid inflationary
scenario for the helicity $1/2$ gravitino states. 
In this case the adiabatic condition is violated maximally near the point
where $m(t)$ is minimal but non zero ( then $|g \phi(t)|$ is maximal),
and this again happens when $\dot \phi(t) \approx 0$. Using
Eq.~(\ref{adia}), we can then easily estimate the upper limit on $|\phi(t)|$
\begin{eqnarray}
\label{estim}
\left(\frac{ m_\phi^2 g \phi(t)}{\sqrt{2}} \right)^{2/3}-(m_X - |g\phi(t)|)^2
\geq k^2\,,
\end{eqnarray}
where we have replaced $\ddot \phi(t) =-m_{\phi}^2\phi(t)$. Notice
that we have  explicitly taken the negative sign for $\phi(t)$, in
order to extremise the left-hand side of the above equation. 
We would like to see the range of the violation of the adiabatic condition 
for small $k$,                            
\begin{eqnarray}
\label{lim}
\frac{m_X}{3} \stackrel{\textstyle{<}}{_\sim} | g \phi(t)|
\stackrel{\textstyle{<}}{_\sim}  m_X \,.
\end{eqnarray}
Hence, in general the production could continue as long as the field amplitude
follows the above condition  Eq.~(\ref{lim}). 

The maximal range of momenta for which the fermions are produced is
obtained when $|g \phi(t)| \approx m_X$, and is given by
\begin{eqnarray}
\label{estim1}
\left(\frac{m_{\rm{\phi}}^2 m_X}{\sqrt{2}}\right)^{1/3} \geq k_{\rm{max}}\,,
\end{eqnarray} 
and for $m_X \approx m_{\rm{\phi}}$, the above expression reduces to
$k_{\rm{max}} \leq m_X$. This result could have been easily 
derived from Eq.~(\ref{con1}) by taking the masses to be almost equal.

In the hybrid model, which we have considered, the amplitude of the
oscillations is small, never exceeding $m_X/g$. Hence we do not expect
to produce super-massive fermions. 
The above result can be directly applied to the production mechanism
of helicity 
$1/2$ gravitinos. We have mentioned in the main section that the
effective time-varying 
mass for the helicity $1/2$ states does not vanish. The effective mass of the
$1/2$ states is almost equal to that of the oscillating frequency, which
corresponds to having $m_{X} \approx m_{\phi}$ in our present discussion.
In addition the adiabatic condition is broken for a narrow range of field values,
see Eq.~(\ref{lim}). This means that the occupation number for the helicity $1/2$
builds up in each and every oscillation and the gravitinos are produced in 
bursts.

 


\begin{references}
\bibitem{nilles} For a review, see H. P. Nilles, Phys. Rept. {\bf 110}, 1
      (1984).
\bibitem{lyth} For a review, see D. H. Lyth and A. Riotto, Phys. Rept.
 {\bf 314}, 1 (1999).
\bibitem{sarkar} For a review, see  S. Sarkar, Rept. Prog. Phys.
     {\bf 59}, 1493  (1996).


\bibitem{linde} L. Kofman, A. Linde and A. Starobinsky, Phys. Rev. Lett.
        {\bf 73}, 3195 (1994); Phys. Rev. D {\bf 56} 3258 (1997);
         Y. Shtanov, J. Traschen and R. H. Brandenberger,
        Phys. Rev. D {\bf 51}, 5438 (1995); D. Boyanovsky, H. J. De Vega,
        R. Holman, D. S. Lee, and A. Singh, Phys. Rev. D {\bf 51}, 4419
        (1995).

\bibitem{patrick}P. Greene and L. Kofman, Phys. Lett. B{\bf 448}, 6 (1999);
            J. Baacke, K. Heitmann and C. Patzold, Phys. Rev. D {\bf58}
        (1998); A. L. Maroto and A. Mazumdar, Phys. Rev. D {\bf 59},
         083510 (1999).

\bibitem{rocky} D. J. H. Chung, E. W. Kolb and A. Riotto, Phys. Rev. D
      {\bf 59}, 023501 (1999); D. J. H. Chung, E. W. Kolb and A. Riotto,
      Phys. Rev. Lett. {\bf 81}, 4048 (1998). 

\bibitem{berezin}For a Review, see V. Berezinsky, astro-ph/9801046.

\bibitem{peloso}G. F. Giudice, M. Peloso, A. Riotto and I. Tkachev,
       JHEP, 9908 014, (1999).

\bibitem{mar}J. Garcia Bellido and A. Linde, Phys. Rev. D {\bf 57},
       6075  (1998).

\bibitem{mar1}M. Bastero-Gil, S. F. King and J. Sanderson, 
       Phys. Rev. D {\bf 60}, 103517 (1999).





\bibitem{velo} G. Velo and D. Zwanziger, Phys. Rev. {\bf 186},
       1337 (1969); C. R. hagen and L. P. S. Singh, Phys. Rev. D
       {\bf 26}, 393 (1982).

\bibitem{deser0} V. P. Akulov, D. V. Volkov and V. A. Soroka, JETP Lett. {\bf 22} 396 (1973); 
S. Deser and B. Zumino, Phys. Lett. B {\bf 62}, 335 (1976)


\bibitem{auvil}P. R. Auvil and J. J. Brehm, Phys. Rev. {\bf 145},
       1152 (1966).

\bibitem{deser} S. Deser and B. Zumino, Phys. Rev. Lett. {\bf 38},
        1433 (1977).

\bibitem{anupam} A. L. maroto and A. Mazumdar, Phys. Rev. Lett. {\bf 84}, 1655 (2000).

\bibitem{ross} G. G. Ross and S. Sarkar, Nucl. Phys. B {\bf 461}, 597
        (1996).

\bibitem{cremmer} E. Cremmer, B. Julia, J. Scherk, S. Ferrera, L. Girardello and P. 
Van Nieuwenhuizen, Nucl. Phys. B {\bf 147}, 105 (1979).


\bibitem{kallosh} R. Kallosh, L. Kofman, A. Linde and A. V. Proeyen,
       Phys. Rev. D {\bf 61},  103503 (2000)

\bibitem{riotto1} G. F. Giudice, I. Tkachev and A. Riotto,
        JHEP 9908:009 (1999).

\bibitem{lyth1} D. H. Lyth, Phys. Lett. B {\bf 469}, 69 (1999);
Phys. Lett. B {\bf 476}, 356 (2000).


\bibitem{riotto2} G. F. Giudice, A. Riotto and I. Tkachev,
       JHEP, 9911, 036 (1999). 

\bibitem{maroto} A. L. Maroto and J. R. Pelaz, Phys. Rev. D {\bf 62}, 023518
 (2000).

\bibitem{hybrid}
 A. D. Linde, Phys. Lett. B {\bf 249}, 18 (1990);
A. D. Linde, Phys. Lett. B {\bf  259}, 38 (1991);   E. J. Copeland,
A. R. Liddle, D. H. Lyth, E. D. Stewart and 
D. Wands, Phys. Rev. B {\bf  49}, 6410 (1994).

\bibitem{baulieu} L. Baulieu, A. Georges and S. Ouvry, Nucl. Phys. B
{\bf 273}, 366 (1986). 

\bibitem{casalbuoni} R. Casalbuoni, S. De Curtis, D. Dominici,
F. Feruglio and R. Gatto, 
Phys. Lett. B {\bf 215}, 313 (1988); Phys. Rev. D {\bf 39}, 2281 (1989).



\bibitem{bailin} D. Bailin and A. Love, {\it Supersymmetric Gauge
   Field Theory and String Theory}, {\bf IOP}, Bristol (1994).

\bibitem{fayet}P. Fayet, Phys. Lett. B {\bf 84}, 421 (1979).


\bibitem{van} P. Van Nieuwenhuizen, Phys. Rep. {\bf 68}, 189 (1981).

\bibitem{bau} L. Baulieu, Phys. Rep. {\bf 129}, 1 (1985).


\bibitem{Dinflation}
 E. Halyo, Phys. Lett. B {\bf  387}, 43 (1996); P. Bin\'etruy and
G. Dvali, Phys. Lett. B {\bf  388}, 241 (1996). 
For earlier work on this subject see: J. A. Casas and C. Mu\~noz,
Phys. Lett. B{\bf  216}, 37 (1989); J. A. Casas, J. Moreno, C. Mu\~noz
 and M. Quiros, Nucl. Phys. B{\bf 328}, 272 (1989).





\bibitem{ma} M. Bastero- Gil and A. Mazumdar, in preparation.
\end{references}
\end{document}